\documentclass[%
preprint,
showkeys,showpacs,
nofootinbib,
nofootinbib,
 amsmath,amssymb,
 aps,
]{revtex4-1}
\usepackage[mathscr]{eucal}
\usepackage{graphicx}
\usepackage{dcolumn}
\usepackage[textwidth=3cm,textsize=tiny,shadow]{todonotes}
\newcommand\abs[1]{\lvert #1 \rvert}
\usepackage{booktabs}

\newtheorem{obs}{Observation}

\begin{document}


\title{Level crossings and new exact solutions of the two-photon Rabi
  model}

\author{Andrzej J. Maciejewski} \email{a.maciejewski@ia.uz.zgora.pl} \affiliation{Uniwersity of Zielona G\'ora, Poland
  }\author{Tomasz Stachowiak}%
\email{stachowiak@cft.edu.pl} \affiliation{%
  Independent Scholar}

\begin{abstract}
  An infinite family of exact solutions of the two-photon Rabi model  
  was found by investigating the differential algebraic properties of
  the Hamiltonian.  This family 
  corresponds to energy level
  crossings not covered by the Juddian class, which is given by elemetary functions.  In contrast, the new states are   expressible in  terms of parabolic cylinder or Bessel
  functions.  We discuss three approaches for discovering this hidden
  structure: factorization of differential equations, Kimura
  transformation, and a doubly-infinite, transcendental basis
  of the Bargmann space.
\end{abstract}

\maketitle


\section{Formulation of the problem}
\label{sec:problem}
Emary and Bishop were the first to describe exact solutions of the
two-photon Rabi model \cite{Emary:02::}.  They related them to
crossings of energy levels with different parities of corresponding
eigenfunctions. These exact solutions are expresible in terms of elementary functions are called the Juddian solutions.  Moreover, Emary and Bishop noticed that their result describes
only a part of the level-crossings which are clearly visible in a
numerically obtained spectrum, and conjectured that the remaining
level-crossings also correspond to exact solutions\footnote{see p. 8240 in~\cite{Emary:02::}.  The authors suggest the the missing exact solutions will be of  Juddian type.}. This conjecture
can be reformulated as the problem of full description of degenerate
states in the 2-photon Rabi model.  Our aim is to give a complete solution
of this problem and to show that Emary and Bishop's conjecture is
true.  We give explicit analytic formulae for eigenfunctions
corresponding to level crossings which were conjectured by Emary and
Bishop.  More importantly, we show that they are of different nature
and analytical properties that those found in~\cite{Emary:02::}. The
proposed method is general, it can be applied to arbitrary systems for
which the Sch\"odringer equation has order greater than two.

The two photon Rabi model is given by the following Hamiltonian
\begin{equation}
  H = \omega a^{\dag}a +\frac{\omega_0}{2}\sigma_z +
  2g \left[(a^{\dag})^2+(a)^2\right]
  \sigma_x.
  \label{H2pp}
\end{equation}
It is also known as the two-photon Jaynes-Cummings model investigated
initially in \cite{Gerry:88::}, and with a view to population inversion
in \cite{Penna:16::}.  This is a phenomenological model of the
two-photon interaction. Its RWA version was introduced
in~\cite{Sukumar:81::}. The reader can find a nice discussion and justification of the derivation in \cite{Emary:01::}.

For very recent studies of the two-photon Rabi model we refer the
reader to \cite{JPA16,JPA17}, where a detailed references for this
subject can be found.

In further consideration it is convenient to apply a unitary
transformation $ \widetilde H= U^\dag H U$ with
$U = (\sigma_x+\sigma_z)/\sqrt2$. It gives
\begin{equation}
  \begin{split}
    \widetilde H&= \omega a^{\dag}a +\frac{\omega_0}{2}\sigma_x +
    2g\left[(a^{\dag})^2+a^2\right]\sigma_z \\
    &= 2g \left\{ 2x a^{\dag}a +\mu\sigma_x +
      \left[(a^{\dag})^2+a^2\right]\sigma_z\right\},
  \end{split}
\end{equation}
where we set $\omega=4xg$, and $\omega_0 =4\mu g$. The rescaled
Hamiltonian in matrix form reads
\begin{equation}
  \label{Kamilton}
  \begin{aligned}
    K&:= \frac{1}{2g} \widetilde{H} = 2x a^{\dag}a +\mu\sigma_x +
    \left[(a^{\dag})^2+a^2\right]\sigma_z \\
    &= \begin{bmatrix}
      2x a^{\dag}a    + \left[(a^{\dag})^2+a^2\right] & \mu \\
      \mu & 2x a^{\dag}a - \left[(a^{\dag})^2+a^2\right]
    \end{bmatrix}.
  \end{aligned}
\end{equation}
We next employ the Bargmann representation, in which the state is a two-component spinor: $\psi\in\mathcal{H}^2$, where $\mathcal{H}$ is the Hilbert space which is a proper $\mathbb{C}$-linear subspace of entire functions $\mathscr{O}(\mathbb{C})$ of the complex variable $z\in\mathbb{C}$, with the scalar product
\begin{equation}
  \langle f,g\rangle=
  \dfrac{1}{\pi}\int_\mathbb{C}\overline{f(z)}g(z)e^{-|z|^2}\mathrm{d}
  (\Re(z))\mathrm{d}(\Im(z)).
\end{equation}
The operators $a^{\dag}$ and $a$ become $z$ and $\partial_z$, respectively, so the stationary Schr\"odinger equation
$K\psi = E\psi$, has the form of the following system of differential
equations
\begin{equation}
  \begin{aligned}
    \psi_1''(z) +2x z\psi_1'(z)+(z^2-E)\psi_1(z)+
    \mu\psi_2(z) &= 0,\\
    \psi_2''(z) -2x z\psi_2'(z)+(z^2+E)\psi_2(z)- \mu\psi_1(z) &= 0.
  \end{aligned}
  \label{eq:sys1}
\end{equation}
In some calculations it is more convenient not to take the above
system, but rather the corresponding fourth-order equation, obtained
by elimination of $\psi_2$,
\begin{multline}
  \label{eq:n=2}
  \varphi^{(\text{iv})} + ((2-4x^2)z^2+4x)\varphi''
  +4z(1+E x -x^2)\varphi'+
  (2-E^2+\mu^2-4xz^2+z^4)\varphi = 0, 
\end{multline}
where we denote $\varphi(z)=\psi_1(z)$; written as $L_{2p}(\varphi)=0$, the equation defines the differential operator $L_{2p}$.
As we showed in \cite{Maciejewski:17::}, the natural parameters of the
problem are $\kappa$ and $\chi$ which are defined by relations
\begin{equation}
  \label{eq:8}
  x=\frac{1}{2}\left(\frac{1}{\kappa}+\kappa\right) \qquad
  E=2\left(\frac{1}{\kappa}-\kappa\right) (\chi-1)-\kappa,
\end{equation}
and $\chi$ is the spectral parameter. Here we assume that
$\kappa\in(0,1)$ and $E>-1$. These restriction of parameters implies
that $\chi\geq 1$. For justification of this  parametrisation  see our
article \cite{Maciejewski:17::}.

For any fixed values of the parameters equation~\eqref{eq:n=2} has four
linearly independent solution $\varphi_i(z)$, $i=1,\ldots, 4$ which
are entire functions.  They are uniquely determined by four linearly
independent initial conditions
$(\varphi_i(z_0),
\varphi_i'(z_0),\varphi_i''(z_0),\varphi_{i}'''(z_0))$, $i=1,\ldots, 4$, given for an
arbitrary $z_0\in\mathbb{C}$. The fundamental problem is to decide if at
least one of these solutions has a finite norm, i.e., whether a fixed
value of $E$ (or $\chi)$ is the eigenvalue for which one or more
solutions of~(\ref{eq:n=2}) are eigenstates.  Thus, potentially, the
degeneracy of a given eigenvalue can be four. However, this bound
cannot be achieved, and we claim that the degeneracy of
eigenvalues is not higher than two.

To show this, let us recall that if an entire function belongs to
Bargmann's Hilbert space, then it has a proper growth order as
$\abs{z}\rightarrow \infty$. This growth is characterized by two
numbers $(\rho,\sigma)$, the order $\rho$ and the type
$\sigma$. Roughly speaking, $f(z)$ has order $\rho$ and type $\sigma$
if $\abs{f(z)}$ behaves as $\exp\left(\sigma \abs{z}^{\rho}\right)$
for large $\abs{z}$, see \cite{Vourdas:06::}. It can be shown that if
$f(z)$ has a finite norm then $\rho\leq2$ and $\sigma\leq 1/2$.

Equation~\eqref{eq:n=2} has an irregular singular point at
infinity. Formal solutions $\widehat\varphi(z)$ near this point give us information about
the asymptotics of all solutions for large $\abs{z}$. The basis
$\widehat\varphi_i(z)$ with $i=1,\ldots,4$ of formal solutions of
equation~\eqref{eq:n=2} is the following:
\begin{equation}
  \label{eq:7}
  \begin{split}
   \widehat\varphi_1(z) = z^{-1/2}\exp \left[\frac{\kappa}{2} z^{2} \right]
    \left( 1+ \mathscr{O}(z^{-2})\right), & \quad\widehat\varphi_2(z) =
    z^{-3/2}\exp \left[-\frac{\kappa}{2} z^{2} \right] \left( 1+
      \mathscr{O}(z^{-2})\right), \\
    \widehat\varphi_3(z)= z^{-9/2}\exp \left[\frac{1}{2\kappa} z^{2}
    \right] \left( 1+ \mathscr{O}(z^{-2})\right), & \quad
    \widehat\varphi_4(z) = z^{-5/2}\exp \left[-\frac{1}{2\kappa} z^{2}
    \right] \left( 1+ \mathscr{O}(z^{-2})\right).
  \end{split}
\end{equation}
The above expansions are asymptotic to true solutions $\varphi_i(z)$
only in a narrow enough sector of the complex plane. From their form
it follows that all solutions of equation~\eqref{eq:n=2} have order
$\rho=2$. However, a changing of the sector permutes these asymptotic
expansions and this is the effect of the Stokes phenomenon, see
\cite{Wasow}. This is why generic solutions have the maximal type
which in our case is $1/2\kappa>1/2$. The above shows that there are
at most two linearly independent solutions of type
$\sigma=\kappa/2<1/2$. For if there existed a solution independent of $\hat\varphi_1$ and $\hat\varphi_2$, but with the type different from $\hat\varphi_3$ and $\hat\varphi_4$, the basis would have 5 elements, by the correspondence between formal and global bases \cite{Wasow}. Thus, we arrive at
\begin{obs}
The degeneracy in the two-photon Rabi model is at most two.
\end{obs}
In any case, the condition that a solution of
equation~\eqref{eq:n=2} is an eigenvector puts restrictions on the
Stokes matrices which are highly transcendental objects.

The two photon Rabi model described by Hamiltonian~\eqref{Kamilton}
has a $\mathbb{Z}_4$ symmetry. The Hamiltonian $K$ commutes with the
operator $\tau$ defined as
\begin{equation}
  \label{eq:tau}
  \tau\psi(z) = \sigma_x \psi(\mathrm{i}z), \qquad \psi=(\psi_1,\psi_2)^T, \quad 
  \sigma_x=\begin{bmatrix}  
    0 &1\\1 &0
  \end{bmatrix}
\end{equation}
The associated symmetry group is $\mathbb{Z}_4$ isomorphic to
$\{1,\tau,\tau^2,\tau^3\}$.

The basis of the solutions space of~\eqref{eq:n=2} has four elements, and they can be numbered according to
their $\mathbb{Z}_4$ parity.
 For a solution with parity $s$ one
has $\tau\psi(z) =s \psi$, so 
\begin{equation}
  \begin{split}
    \psi_1(\mathrm{i} z) &= s \psi_2(z),\\
    \psi_2(\mathrm{i} z) &= s \psi_1(z),
  \end{split}
\end{equation}
where $s\in\{+1,-1,\mathrm{i},-\mathrm{i}\}$. Because
$\tau^2\psi(z) = \psi(-z)$, the distinction between even and odd
solutions represents the $\mathbb{Z}_2$ subgroup of the symmetry and
can be used to simplify the calculation.

We can distinguish four linearly independent solutions of system~\eqref{eq:n=2}:  
$\psi^{(s)}=(\psi_1^{(s)}, \psi_2^{(s)})$ with
$s\in\{+1,-1,\mathrm{i},-\mathrm{i}\}$ just specifying initial
conditions $(\psi^{(s)}(0), {\psi^{(s)}}'(0))$. A simple reasoning
shows that if $s\in\{-1,+1\}$ then $\psi^{(s)}(0)=(1,s)$ and
$ {\psi^{(s)}}'(0)=(0,0)$.  In this case
$\psi^{(s)}(z)=\psi^{(s)}(-z)$.  If $s\in\{\mathrm{i},-\mathrm{i}\}$,
then $\psi^{(s)}(0)=(0,0)$ and $ {\psi^{(s)}}'(0)=(1,-\mathrm{i}s)$;
in this case $\psi^{(s)}(z)=-\psi^{(s)}(-z)$. Clearly, these four initial
conditions are linearly independent. Each solution $\psi(z)$ of system~\eqref{eq:n=2}
is a linear combination
\begin{equation}
\label{eq:1}
\psi(z)= c_{-1}\psi^{(-1)}(z)+ c_{+1}\psi^{(+1)}(z)+ c_{-\mathrm{i}}\psi^{(-\mathrm{i})}(z)+ c_{+\mathrm{i}}\psi^{(+\mathrm{i})}(z).
\end{equation}
In other words, for a specific energy $E$, once the parity $s$ is fixed, so are the initial conditions at $z=0$, and the state is uniquely determined up to a multiplicative constant.

\begin{figure}[h]
  \includegraphics[scale=0.43]{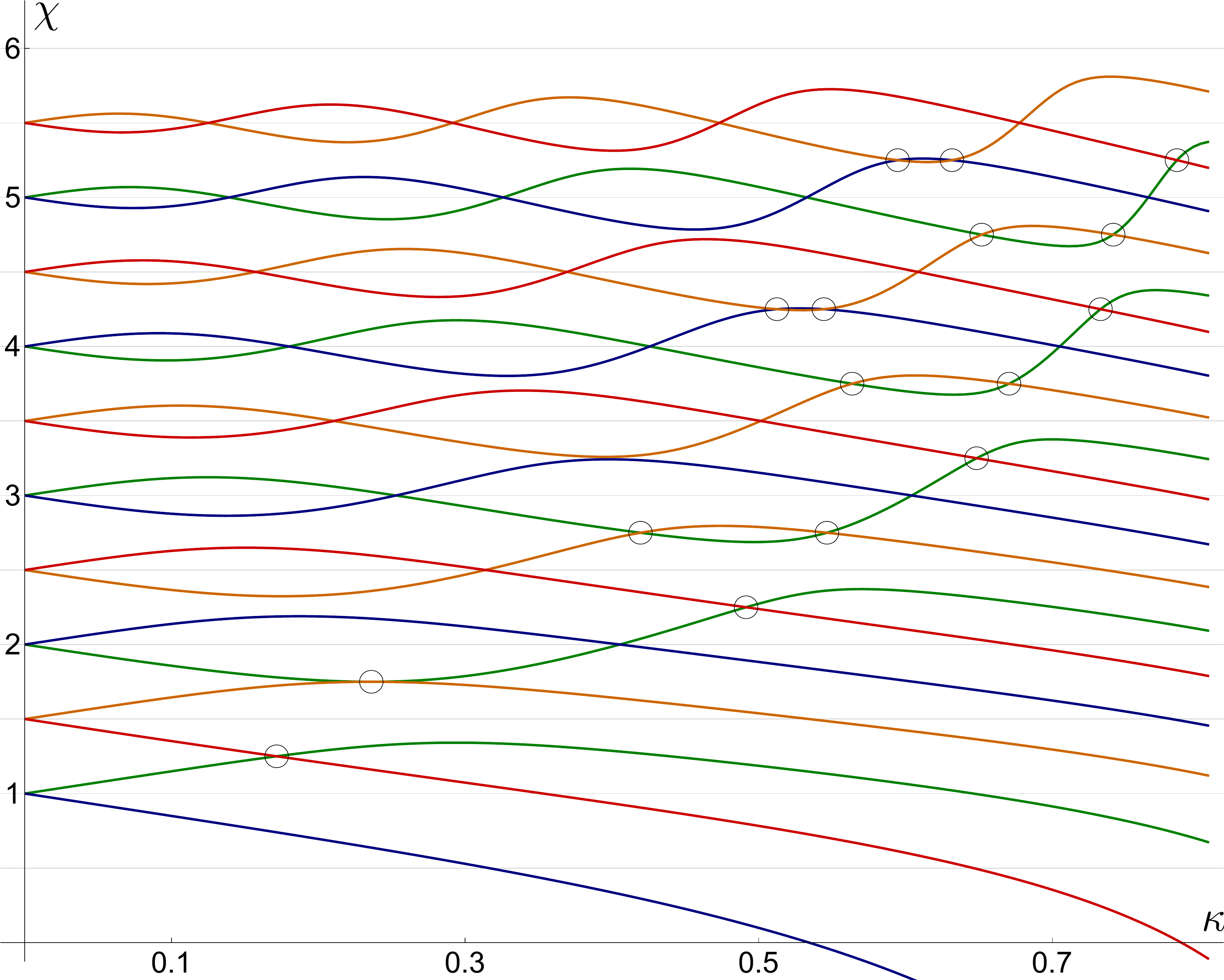}
  \caption{\label{fig:spec} Spectrum of the two-photon Rabi model for
    $\mu=3$, parities: blue: $-1$, green: $+1$, red: $-\text{i}$, orange: $+\text{i}$. Circles indicate the new states; the remaining crossings are Juddian.}
\end{figure}

The spectrum of the two-photon Rabi model is shown in
Figure~\ref{fig:spec}.  The color of a curve marks the parity of the
corresponding eigenfunction: green, blue, red and orange for $+1$,
$-1$, $+\mathrm{i}$ and $-\mathrm{i}$, respectively.  The Judd states
correspond to crossings at levels $\chi=n/2$, with integer $n>1$. They
occur only as intersections of two curves with parities $(+1,-1)$ or
$(+\mathrm{i}, -\mathrm{i})$. The intersections marked with circles
correspond to new states discussed in this papers.  For them there are
four admissibly parities of intersecting curves
$(\pm 1, \pm\mathrm{i})$. Values of $\chi$ for these crossings are
$\chi=(2\ell+3)/4$ for integer $\ell\geq 1$. This fact is not trivial as the
curves are know only numerically. In fact, at first we determined
these values of $\chi$ investing continued fractions of successive
numerical approximations. Later we show how these distinguished values
of $\chi$ naturally appear in our approach.
In the next section we describe the main idea of our approach to the
problem.  The eigenvectors corresponding to special points of the spectrum
are distinguished by their analytical form: for Judd states they are given by elementary function while for the remaining ones -- by higher transcendental functions (Bessel, or parabolic cylinder). The reason why this
simplification occurs is connected to an algebraic property of the
differential operator defined by the stationary Schr\"odinger
equation. Physical manifestation of this property is the degeneracy of the spectrum, and the main result can be summarized as follows:
\begin{obs}
Degeneracy in the two-photon Rabi model occurs when $\chi=m/4$, with $m\in\mathbb{Z}_+$, and the Schroedinger equation is factorizable. Juddian solutions appear for  even $m$, and in this case  the equation has a 1st order factor; for odd $m=2\ell +3$,  $\ell\in\mathbb{Z}_+$, the minimal factor is of 2nd order, and for each $\ell$ the degenerate eigenstates are given by classical transcendental functions by formula~\eqref{eq:32};  the appropriate algebraic conditions on $\kappa$ and $\mu$ is given by the determinant of equations~\eqref{eq:333}.
\end{obs}

\section{Factorization}

\subsection{First approach}
\label{sec:a1}

The exact solutions of the two-photon Rabi model found by Emary and
Bishop ~\cite{Emary:02::} correspond to $\chi=n/2$ , where $n$ is an
integer, and are directly related to the Judd solutions of the standard
Rabi model \cite{Emary:02::} in two ways. First, the solutions are
exponential
\begin{equation}
  \psi_1(z) = \mathrm{e}^{a z^2} P(z),
\end{equation}
where $P(z)$ is a polynomial, and $2a\in\{\pm\kappa,\pm\kappa^{-1}\}$.
Second, they come about by
degeneracy of energy levels with different parities. 
More precisely,  the eigenspace of a Judd state  is two dimensional and all eigenvectors $\psi(z)=(\psi_1(z),\psi_2(z))$  from this space are either even, when it appears as a level crossing of states with $\mathbb{Z}_4$ parities $+1$ and $-1$, or they are odd -- for level crossing of states with parities , $+\mathrm{i}$ and $-\mathrm{i}$.

Moreover, from the mathematical point of view, what happens is that
the differential equation~\eqref{eq:n=2} is reducible, i.e., the
corresponding differential operator $L_{\mathrm{2p}}$ can be
factorized
\begin{equation}
  L_{\mathrm{2p}}= L_3 L_1.
\end{equation}
where $L_1$ is a first order differential operator. It seems that this fact was never mentioned in the literature in this context.

Knowing the above  we asked if it is possible that the operator
$L_{\mathrm{2p}}$ has a right factor of second order, and if so -- what is the physical meaning of this kind of factorization?

Let us notice that if $L_{\mathrm{2p}}$ admits the factorization
\begin{equation}
  L_{\mathrm{2p}}=\widetilde L_2 L_2,
\end{equation}
then a solution of equation $L_2(y)=0$ is also a solution
equation~\eqref{eq:n=2}. Thus, the question is if there exist energies
for which states are solutions of second order differential equations. 
We take the factor to be minimal, for if $L_2=\tilde{L}_1 L_1$, we would be back in the previous case. In the mathematical literature functions which are solutions  of a second order linear differential equations are called Eulerian, see \cite{Singer85}.

The problem of factorization of differential operator is a classical
subject. A short exposition can be found
in~\cite{Bronstein:94::}. Using the algorithm and facts from this reference
it can be shown that:
\begin{enumerate}
\item\label{item:1} if the differential operator $L_{\mathrm{2p}}$ has a
  right second-order factor $L_2$, then
  \begin{equation}
    \label{eq:9}
    \chi=\frac{2\ell+3}{4}, \qquad \ell\in \mathbb{Z};
  \end{equation}
\item the remaining parameters satisfy the algebraic condition
  $P_\ell(\mu,\kappa)P_\ell(-\mu,\kappa)=0$.  where
  $P_\ell(x,y)\in\mathbb{Z}[x,y]$ is a polynomial with integer
  coefficients;
\item for $\ell>0$ \textbf{all solutions} of the right factor
  $L_2 (\psi_1)=0$ have a finite Bargmann norm; for $\ell\leq 0$ these
  solutions are not normalizable;
\end{enumerate}
For $\ell=1$, i.e., for $\chi=5/4 $, the right factor $L_2$ of
$L_{\mathrm{2p}}$ is given by
\begin{equation}
  \label{eq:eq54} 
  L_2 = \partial_z^2 +\kappa(2-\kappa z^2 ),  
\end{equation}
The algebraic restriction for the other parameters is given by the polynomial
$P_1(\mu,\kappa)=(2\mu \kappa+\kappa^2+1) $. It is important to remark
here that an arbitrary solution $V(z)$ of the equation $L_2(V)=0$ has
finite norm.  In fact, the equation
\begin{equation}
  \label{eq:10}
  L_2(V)= V''(z) +\kappa(2-\kappa z^2 ) V(z) =0 
\end{equation}
has only one irregular singular point at infinity so all its solutions
are entire functions.  Moreover, the basis of formal solutions near
infinity is
\begin{equation}
  \label{eq:11}
  \widehat V_1(z) = z^{1/2}\mathrm{e}^{-\frac{\kappa}{2}z^2 } \left( 1 +O(z^{-2}) \right), \quad
  \widehat V_2(z) = z^{-3/2}\mathrm{e}^{\frac{\kappa}{2}z^2 } \left( 1 +O(z^{-2}) \right), \quad
\end{equation}
An entire function $V(z)$ with these asymptotics has order $2$ and
type $\kappa/2$. As $\kappa\in(0,1)$, the Bargmann norm of $V(z)$ is
finite. Two linearly independent solution of~\eqref{eq:10} are 
\begin{equation}
    \label{eq:pc}
    V_1(z)= D_{1/2}(\sqrt{2\kappa}z),  \qquad  V_2(z)=D_{1/2}(-\sqrt{2\kappa}z)
\end{equation}
where $D_n(z)$ denotes the parabolic cylinder function in Whittaker's convention, see \cite[\S 16.5]{Whittaker:35::}; for a compilation of definitions, conventions and properties see \cite{DLMF}. These two functions  form a basis of the eigenspace    for $\chi=5/4 $, i.e., 
\[
\psi_1(z)= c_1 D_{1/2}(\sqrt{2\kappa}z) +c_2 D_{1/2}(-\sqrt{2\kappa}z)
\]
where $c_1$ and $c_2$ are arbitrary constants.

For $\ell=2$ the right factor of $L_{2\mathrm{p}}$ is
\begin{equation}
  \label{eq:eq74}
  L_2 = \partial_z^2 +b_1(z)\partial_z + b_0,  
\end{equation}
where
\begin{equation}
  b_1(z)= -\frac{2(1+\kappa^2)z}{(1+\kappa^2)z^2 -2\kappa}  
\end{equation}
\begin{equation}
  b_0(z)= 2 \kappa  - \kappa^2 z^2 +\frac{ 9-2(11+2\mu^2)\kappa^2 +
    \kappa^4}{4(1-\kappa^2)\left[(1+\kappa^2)z^2 -2\kappa\right]}.  
\end{equation}
The algebraic restrictions are given by the polynomial
\begin{equation}
  P_2(\mu,\kappa) =3(1+\kappa^2)^2 - 8\mu\kappa(1-\kappa^2) +
  4   \mu^2\kappa^2.
\end{equation}
Let us notice that now the differential equation $L_2(V)=0$ has regular
singularities at points
\begin{equation}
  \label{eq:12}
  z_{\pm}=\pm \sqrt{\frac{2\kappa}{1+\kappa^{2}}}.
\end{equation}
The differences of exponents at these points are both equal 2 but
nevertheless all its solutions are entire. The algebraic condition
$P_2(\mu,\kappa) P_{2}(-\mu,\kappa)=0$ guarantees 
a that $L_2$ is the right factor of $L_{2\mathrm{p}}$, and this in
turn implies the logarithmic therms are not present in the solutions
of $L_2(V)=0$.  Quite amazingly, the basis of formal solution of
$L_2(V)=0$ near infinity is the same as for the respective equation
for $\ell=1$, that is it has the form~\eqref{eq:11}.
We postpone giving explicit solutions until the next two section, where different values of $\ell$ can be treated together.

Taking $\ell>2$ we obtain the right factor $L_2$ which is of the
form~\eqref{eq:eq74} with rational $b_1(z)$ and $b_0(z)$. However, the
number of regular singularities is $2(\ell-1)$, and the expressions obtained 
for the coefficients are quite complicated. For the regular singularities
the differences of exponents are integers. However, the condition
$P_{\ell}(\mu,\kappa) P_{\ell}(-\mu,\kappa)=0$ guarantees that all these
singularity are apparent, see \cite{Barkatou:15::}. Thus, one would
like to remove them by transforming the factor equation $L_2(V)=0$ into
a second order equation without singular points in the finite part of
the complex plane.  This question reformulated in more general
settings is not new. From the relevant theory \cite{Barkatou:15::} it follows that the answer to
this question is generally negative, and known methods are not directly
applicable for our problem.  This is why we decided to modify our
approach.

\subsection{Second approach}
\label{sec:a2}
Equation~\eqref{eq:n=2} has a second order factor if its solution
space has a two dimensional subspace spanned by a basis of solutions
of second order equation.  The real problem is the dependence on
parameters. We know that necessary condition for the factorization is
$\ell\in\mathbb{Z}$. Let $\mathscr{W}_\ell$ denote the above
mentioned subspace of solutions for $\ell>0$. Our crucial observation
is that for each $\ell\in\mathbb{Z}$ if $V(z)\in \mathscr{W}_\ell$,
then
\begin{equation}
  \label{eq:13}
  V(z) = A_\ell(z) d(z) + B_\ell(z) d'(z),
\end{equation} 
where $A_\ell(z)$, $B_\ell(z)$ are polynomials, and $d(z)$ is an
arbitrary solution of the following differential equation
\begin{equation}
  \label{eq:14}
  d''(z)+ \kappa(2 - \kappa z^2)d(z)=0.
\end{equation}
This kind of transformation often appears in the problem of removing
apparent singularities, as introduced by Kimura in application to the
hypergeometric equation \cite{Kimura:70::}.

Here it is important to underline that
\begin{enumerate}
\item $d(z)$ does not depend on $\ell$;
\item an arbitrary solution $d(z)$ of \eqref{eq:14} has finite
  Bargmann norm (see below), so all elements of $\mathscr{W}_\ell$
  have finite Bargmann norm;
\item the function $V(z)$ given by~\eqref{eq:13} is a solution of
  ~\eqref{eq:n=2} if and only if the condition
  $P_\ell(\mu,\kappa) P_\ell(-\mu,\kappa)=0$ is satisfied;
\item the polynomial $A_\ell(z)$ has degree $(\ell-1)$ and
  $A_\ell(-z)=(-1)^{\ell+1}A_\ell(z)$;
\item the polynomial $B_\ell(z)$ has degree $(\ell-2)$, $B_2(z)=0$ and
  $B_\ell(-z)=(-1)^{\ell}B_\ell(z)$.
\end{enumerate}
Crucially, from the practical point of view, it is enough to assume
that $\psi_1(z)= A(z)d(z) +B(z)d'(z)$, where $A(z)$ and $B(z)$ are
polynomials, is a solution of~\eqref{eq:n=2}, and we can derive all
the conclusions given above.

In the preceding section, the factors $L_2$ depended on $\ell$. Here,
there is just the equation for $d(z)$ and all solutions will involve
only two transcendental functions, i.e., the two independent solutions
of \eqref{eq:14}. Recall that the classical Judd states are given by
polynomials and exponentials -- the latter being a solution of a first
order differential equation, this means the solutions are
Liouvillian. Here, by contrast, in addition to polynomials we have
solutions of a second-order equation, that is to say: Eulerian
functions. In this case they are the modified Bessel functions of the
first kind, for we have
\begin{equation}
  d^{(\pm)}(z) =
  z^{\frac32}I_{\pm\frac14}\left(\frac{\kappa}{2}z^2\right)
  - z^{\frac32}I_{\mp\frac34}\left(\frac{\kappa}{2}z^2\right).
  \label{besle}
\end{equation}
The algebraic singularity is apparent, since the equation only admits
entire solutions -- the above can be analytically continued over the
whole complex plane. The above form of solution appears  naturally when the Laplace transformation is used  for finding form~\eqref{eq:13}. The basis $d^{\pm}(z)$ has prescribed  symmetry as the notation suggests. Namely, equation~\eqref{eq:14} coincides with~\eqref{eq:10} and one can check that  
\begin{equation}
\label{eq:15}
d^{\pm}(z)= c_{\pm} \left[  D_{1/2}(\sqrt{2\kappa}z) \pm D_{1/2}(-\sqrt{2\kappa}z) \right]
\end{equation}
where $c_{\pm}$ are non-zero constants.   
In connection
with the results of Dolya \cite{Dolya}, one more representation of the
general solution of \eqref{eq:14} would be through the confluent
hypergeometric function:
\begin{equation}
  d(z) = c_1 \text{e}^{-\frac{\kappa}{2}z^2}
  {}_1F_1\left(-\tfrac14,\tfrac12,\kappa z^2\right) +
  c_2 z \text{e}^{-\frac{\kappa}{2}z^2}
  {}_1F_1\left(\tfrac14,\tfrac32,\kappa z^2\right).
\end{equation}

For $\ell=1$ the solution becomes just
\begin{equation}
  \psi_1 = d(z).
\end{equation}

For $\ell=2$ we have
\begin{equation}
  \psi_1 = \kappa\left(7-\kappa^4+\kappa^2(6+4\mu^2)\right) z d(z) + 
  \left(7\kappa^4+\kappa^2(6+4\mu^2)-1\right) d'(z).
\end{equation}

For $\ell=3$ we have
\begin{equation}
\begin{aligned}
   \psi_1 = &\left(
   z^2 +\tfrac{519\kappa^8-12(159+2\mu^2)\kappa^6-2(1155+8\mu^2(31+\mu^2))\kappa^4+4(27+10\mu^2)\kappa^2-9}{256\kappa(1-\kappa^4)(1-\kappa^2)^2}
   \right) d(z) +\\
   &\tfrac{\kappa^2(62+4\mu^2)-1-\kappa^4)}
   {16\kappa(1-\kappa^2)^2}z d'(z).
\end{aligned}
\end{equation}
We deduced representation~\eqref{eq:13} partly in the process of removing apparent singularities with Kimura's method, and partly by inspecting several examples.  Clearly, it needs further mathematical investigation, which we attempt below.

\subsection{Third approach}
\label{sec:a3}

The third and final picture of the new solutions is a simplification
of the previous representation, which uses one special function $d(z)$
and polynomial coefficients. Instead of finding coefficients of
polynomials, as solutions of a differential equation, one might wish to
construct a new basis  in the Hilbert space, such that the
solutions are simply finite linear combinations, rendering the problem
equivalent to solving a linear system. To this end we introduce a new
independent variable $\zeta =\sqrt{2\kappa}z$, and consider
\eqref{eq:14} generalized thus
\begin{equation}
  \mathscr{D}_n''(\zeta) + 
  \left(n+1-\tfrac14\zeta^2\right)\mathscr{D}_n(\zeta) = 0.
\end{equation}
The previously used function corresponds to $n=0$ so that
$d(z)=\mathscr{D}_0(\sqrt{2\kappa}z)$, and it is clear that $\mathscr{D}_n$ are entire
functions for all $n$. The two independent solutions can be taken as
$\mathscr{D}^{(\pm)}_n(\zeta):=D_{n+\frac12}(\zeta)\pm D_{n+\frac12}(-\zeta)$,
where $D_n$ are the aforementioned parabolic cylindrical functions. In addition to the differential equation, the new
functions satisfy the relations 
\begin{equation}
    \tfrac12\zeta \mathscr{D}_n - \mathscr{D}_n' = \mathscr{D}_{n+1},\quad\text{and}\quad
    \tfrac12\zeta \mathscr{D}_n + \mathscr{D}_n' = (n+\tfrac12)\mathscr{D}_{n-1},
\end{equation}
so it makes sense to introduce the raising and lowering operators
\begin{equation}
  \left.\begin{aligned}
    b^{\dag}&:=\tfrac12\zeta-\partial_{\zeta}\\
    b &:=\tfrac12\zeta+\partial_{\zeta}
  \end{aligned} \,\right\} \quad\Longrightarrow \quad
  \left\{\begin{aligned}
  b^{\dag}\mathscr{D}_n &=\mathscr{D}_{n+1}\\
  b\mathscr{D}_n &= \left(n+\tfrac12\right)\mathscr{D}_{n-1}
  \end{aligned}\right.,
  \label{D_rels}
\end{equation}
which obey $[b,b^{\dag}]=1$. We use the conventional notation, but note that they are not Hermitian conjugates in the Bargmann product. 
What is more, unlike in the standard case, $\mathscr{D}_0$ is not annihilated by $b$, so we have a doubly infinite series, i.e., $n\in\mathbb{Z}$.

The triple $b^{\dag}b$, $b^2$ and ${b^{\dag}}^2$ satisfy the commutation relations of the $\frak{su}(1,1)$ algebra
\begin{equation}
    \left[b^{\dag}b+\tfrac12,{b^{\dag}}^2\right] = 2{b^{\dag}}^2,\quad
    \left[b^{\dag}b+\tfrac12,b^2\right] = -2b^2,\quad
    \left[b^2,{b^{\dag}}^2\right] = 4\left(b^{\dag}b+\tfrac12\right).
\end{equation}
However, we cannot immediately choose one of the irreducible representations because here $b^{\dag}$ does not denote the Hermitian conjugate of $b$.  A further Lie algebraic analysis will be given in a  separate publication, Here, we follow the direct path, which is simple enough in this case.

The Schr\"odinger equation \eqref{eq:sys1} can now be written as
\begin{equation}
  \begin{aligned}
    N\psi_1 + \frac{1+\kappa^2}{1-\kappa^2}b^2\psi_1
    +(2-2\chi)\psi_1+\frac{\mu\kappa}{1-\kappa^2}\psi_2 &=0,\\
    N\psi_2 + \frac{1+\kappa^2}{1-\kappa^2}{b^{\dag}}^2\psi_2
    -(1-2\chi)\psi_2-\frac{\mu\kappa}{1-\kappa^2}\psi_1 &=0,\\
  \end{aligned}
  \label{D_sch}
\end{equation}
where $N:=b^{\dag}b$ so that $N\mathscr{D}_n=\left(n+\frac12\right)\mathscr{D}_n$.  As can
be seen, all the involved operators, when applied to $\mathscr{D}_n$, shift the
index by 0 or 2. We can thus look for solutions of the form
\begin{equation}
  \psi_1 = \sum_{j=0}^{l_3}a_{j}\mathscr{D}_{l_1+2j},\quad
  \psi_2 = \sum_{j=0}^{l_4}c_{j}\mathscr{D}_{l_2+2j},
  \label{eq:32}
\end{equation}
i.e., finite expansions in terms of entire functions like in \eqref{eq:13}, except this time the coefficients are constant, and there is more than one function $d(z)$.

The new parameters can be limited, by inspecting the terms with the
lowest indices: in the first component of \eqref{D_sch}, $b^2\psi_1$
must cancel with $\psi_2$; in the second, ${b^{\dag}}^2\psi_2$ must
cancel with $\psi_1$. The two conditions are $l_1-2=l_2$, and
$l_2+2l_4+2 = l_1+2l_3$, or simply: $l_4=l_3$ and $l_2=l_1-2$. It then follows
that the highest terms in the first equation come from
$N\psi_1+(2-2\chi)\psi_1$, and the lowest in the second from
$N\psi_2-(1-2\chi)\psi_2$, their cancellation requires that
\begin{equation}
  \chi = \frac{3+2l_3}{4}, \quad\text{and}\quad l_1 = -1-l_3.
\end{equation}
By \eqref{eq:9}, we recognize that $l_3=\ell$, so $\chi$ fully specifies the index ranges.

We are now ready to turn the Schr\"odinger equation into a homogeneous linear systems of equations
for $a_j$ and $c_j$, with a tridiagonal structure. Indeed, the first equation in \eqref{D_sch} is easily solved for $\psi_2$, and direct substitution changes the second one into $\mathscr{L}\psi_1=0$,
where the operator $\mathscr{L}(N,b^2,{b^{\dag}}^2)$ acts on $\mathscr{D}_n$
simply as
$\mathscr{L}\mathscr{D}_n = \beta_n \mathscr{D}_n+\alpha_n \mathscr{D}_{n+2}+\gamma_{n}\mathscr{D}_{n-2}$ where
\begin{equation}
\begin{aligned}
    \alpha_n &= (1-\kappa^4)(5+2n-4\chi),\\
    \beta_n &= (1+\kappa^4)(4n^2-1)+2(1-\kappa^2)^2(2n-1
    -2\chi(2\chi-3)) + 2\kappa^2\mu^2,\\
    \gamma_n & = (1-\kappa^4)(n^2-\tfrac14)(4\chi+2n-5).
\end{aligned}
\end{equation}
With the substitution \eqref{eq:32}, the remaining equation becomes $\mathscr{L}\sum a_n\mathscr{D}_{l_1+2n}=0$, or, by independence of $\mathscr{D}_n$, a set of recurrence relations
\begin{equation}
    \alpha_{l_1+2n-2}a_{n-1} + \beta_{l_1+2n}a_{n} +
    \gamma_{l_1+2n+2}a_{n+1} = 0,\qquad n=0,\ldots,\ell;
\end{equation}
the equivalent system reads

\begin{equation}
\label{eq:333}
  \begin{bmatrix} 
    \beta_{-\ell-1} & \gamma_{-\ell+1} & & & &\\
    \alpha_{-\ell-1} & \beta_{-\ell+1} & \gamma_{-\ell+3} & 
    & \scalebox{1.5}{0}&\\
    & \alpha_{-\ell+1} & \ddots & \ddots & & \\
    & & \ddots & \ddots & \gamma_{\ell-3}& \\
    & \scalebox{1.5}{0}& & \alpha_{\ell-5} & \beta_{\ell-3} & \gamma_{\ell-1}\\
    & & & & \alpha_{\ell-3} & \beta_{\ell-1}
  \end{bmatrix}
  \begin{bmatrix}
    a_0\\a_1\\ a_2 \\ \vdots \\ a_{\ell-1}\\ a_{\ell}
  \end{bmatrix} =0,
\end{equation}
which is of finite size $\ell+1$ by assumption. For a non-zero solution, the main determinant must vanish -- it turns out to be exactly the
algebraic condition $P_\ell(\mu,\kappa)P_\ell(-\mu,\kappa)=0$. Since $\mathscr{D}_n$ could be any of the
basis solutions, we have obtained solutions for both parities: as
$\tau^2\psi(z)=\psi(-z)=s^2\psi(z)$, the parity of $\mathscr{D}^{(+)}$ is in
$\{1,-1\}$, while that of $\mathscr{D}^{(-)}$ belongs to
$\{\mathrm{i},-\mathrm{i}\}$. We thus have ``cross'' degeneracy. To
determine $s$ specifically one can use the connection formula
\begin{equation}
  \mathscr{D}_n(\mathrm{i}\zeta) = \frac{\mathrm{i}\,
    \Gamma\!\left(n+\tfrac32\right)}{\sqrt{2\pi}}
  \left[\mathrm{i}^{n-\frac12} \mathscr{D}_{-n-2}(-\zeta)
    -\mathrm{i}^{\frac12-n}\mathscr{D}_{-n-2}(\zeta)\right],
\end{equation}
and check the quantity $\psi_1(\text{i}z)/\psi_2(z)$. Another option is to check the expansion of $\psi$ around zero, once the coefficients of \eqref{eq:32} have been solved for, and compare with those in Section~{\bf I}.

The full solutions for $\chi=\frac54$ are
\begin{equation}
  \psi_1 = \mathscr{D}_0,\quad \psi_2 = \frac{1+\kappa^2}{\mu\kappa}\mathscr{D}_{-2},
\end{equation}
where $\mathscr{D}$ of the same parity should be taken in both components.

Likewise, solutions for $\chi=\frac74$ read
\begin{equation}
  \begin{aligned}
    \psi_1 &= \frac{8\kappa^2(1-\kappa^2)\mu^2}{(1+\kappa^2)(3+3\kappa^4+\kappa^2(6+4\mu^2))}\mathscr{D}_{-1}+\mathscr{D}_{1},\\
    \psi_2 &=
    \frac{6\kappa(\kappa^2-1)\mu}{3+3\kappa^4+\kappa^2(6+4\mu^2)}\mathscr{D}_{-3}+\frac{\kappa\mu}{1+\kappa^2}\mathscr{D}_{-1}.
  \end{aligned}
\end{equation}

Getting back to the previous representation is also straightforward,
due to the recurrence relation
\begin{equation}
  \zeta \mathscr{D}_n = \left(n+\tfrac12\right)\mathscr{D}_{n-1}+\mathscr{D}_{n+1},
\end{equation}
which allows us to reduce all $\mathscr{D}_n$ to a combination of $\mathscr{D}_0$ and
$\mathscr{D}_1$. This comes at the price of generating polynomials of $\zeta$ as
coefficients. $\mathscr{D}_1$ can then be changed to $\mathscr{D}_0'$ using
\eqref{D_rels}, and we are back to the $V=A d+ B d'$ form.

By extension, the recurrence relations allow us to manipulate the Bessel functions of \eqref{besle}, and the solutions can be expressed as a sum of $I_\nu$ of quarter order $\nu\in\frac14+\mathbb{Z}$. This is another reason for adoption of $\chi$ as a natural parameter -- quarters correspond to the newly found states, while half integer values (for which the Bessel function reduce to exponentials) correspond to those of Emary and Bishop. The relation of Juddian states to elementary Bessel functions was also the basis of Reik's approach \cite{Reik}.

\section{Symmetry and degeneracy}
As we already mentioned, the Hamiltonian $K$ of the two-photon Rabi model
commutes with the symmetry operator $\tau$ defined by~\eqref{eq:tau}.
Thus, if $H\psi=E\psi$, then  $H\tau(\psi)=E\tau(\psi)$. Hence
\begin{itemize}
\item if $\psi$ is an eigenstate for given energy than also  $\tau(\psi)$,  $\tau^2(\psi)$,  $\tau^{3}(\psi)$, are eigenstate for this energy.
\item if $\mathscr{W}_{E}$  is eigenspace corresponding to energy $E$ then it is spanned by eigenvectors of $\tau$
\end{itemize}
In particular, if energy  $E$ is not degenerate, then necessarily  $\tau(\psi)=s\psi$   with  $s\in\{+1,-1,\mathrm{i},-\mathrm{i}\}$. In other words, non-degenerate eigenstates have specified parity with respect to $\mathbb{Z}_4$ symmetry.

If energy $E$ is degenerate, then we know that $\mathscr{W}_{E}$ has dimension
two, so it is spanned by two eigenstates $\psi_{(r)}$ and
$\psi_{(s)}$ with $r, s\in\{+1,-1,\mathrm{i},-\mathrm{i}\}$,
$r\neq s$, which satisfy $\tau(\psi_{(r)})=r \psi_{(r)}$,
$\tau(\psi_{(s)})=s \psi_{(s)}$.  There are  four unordered  pairs $(r,s)$ of parities
$(1,-1)$, $(1,\mathrm{i})$, $(1,-\mathrm{i})$, and $(\mathrm{i},-\mathrm{i})$.

The Judd type states  found by Emary and Bishop  correspond to
pairs  $(r,s)=(1,-1)$ and  $(\mathrm{i},-\mathrm{i})$. In the first case
$\mathscr{W}_{E}$ is spanned by states which are even functions, while in the second case
$\mathscr{W}_{E}$ is spanned by odd states. Thus the Judd type states of  Emary and Bishop
have fixed $\mathbb{Z}_{2}$ parities corresponding to $\tau^2$.

\section{Conclusions and remarks}
In this article we investigated degenerate states of the two-photon
Rabi model. One part of the collection of such states, called the Judd type states, was
discovered by Emary and Bishop in~\cite{Emary:02::}. The spinors are given by
simple elementary functions.  We observe that this fact is a
consequence of factorisation of the stationary Schr\"odinger equation,
which, in Bargmann's representation, has the form of a fourth order
differential equations. The existence of the other part of the degenerate spectrum was a conjecture~\cite{Emary:02::}. We proved this conjecture, and found the analytical form of the eigenstates and energies.

Our main idea was that degenerate eigenstates have simpler analytical
form than those nondegenerate, because they are the solutions of a
differential equation of order lower than the order of the
Schr\"odinger equation. We determined the asymptotics of the solutions,
and showed that the degeneracy can at most be two-fold. The resulting two-dimensional eigenspaces are spanned by states with two
different $\mathbb{Z}_4$ parities.

The existence of such a lower-order equation in those subspaces means that the  Schr\"odinger
equation considered as a differential operator has a right factor.  We
showed that the remaining part of the degenerate spectrum corresponds to
the case when this factor has order two, so the eigenstates are given
explicitly by solutions of second order differential equations and
higher transcendental functions (equivalently by parabolic cylinder,
Bessel or confluent hypergeometric).

The factorisation method for differential operators is well known in
the context of quantum mechanics, see eg. book~\cite{Dong:07::}.
However, standard techniques consist of factorisation of the
Hamiltonian -- here, in contrast, we factorised the operator corresponding to the whole Schr\"odinger equation.

In the article~\cite{Dolya}, partially overlapping with our analysis, Dolya investigated quasiexact solvability of
the two-photon Rabi system, and obtained similar expressions for eigenstates given
by confluent hypergeometric functions, however this result was
discarded in~\cite{Yao:13::}.

\end{document}